\documentclass[prb,aps,amsmath,amssymb,amsfonts,floatfix,superscriptaddress,showpacs,twocolumn,nofootinbib]{revtex4-1}
\usepackage{graphicx}
\usepackage{color}
\usepackage{bbm}
\usepackage[utf8]{inputenc}
\usepackage{dcolumn}
\usepackage{hyperref}
\usepackage{textcomp}

\usepackage{comment}

\hypersetup{colorlinks=true,breaklinks,linkcolor=blue,urlcolor=blue,citecolor=blue}

\newcommand{\kv}{\ensuremath{\mathbf{k}}}
\newcommand{\av}[1]{\ensuremath{\left\langle #1 \right\rangle}}
\newcommand{\up}{\ensuremath{\uparrow}}
\newcommand{\dn}{\ensuremath{\downarrow}}

\newcommand{\chiimp}{\ensuremath{\chi^{\text{imp}}}}
\newcommand{\Xch}{\ensuremath{X^{\text{ch}}}}
\newcommand{\Xsz}{\ensuremath{X^{\text{sz}}}}

\begin{document}

\title{Double occupancy in DMFT and the Dual Boson approach}

\author{Erik G. C. P. van Loon}
\affiliation{Radboud University, Institute for Molecules and Materials, NL-6525 AJ Nijmegen, The Netherlands}
\author{Friedrich Krien}
\affiliation{Institute of Theoretical Physics, University of Hamburg, 20355 Hamburg, Germany}
\author{Hartmut Hafermann}
\affiliation{Mathematical and Algorithmic Sciences Lab, France Research Center, Huawei Technologies Co. Ltd., 92100 Boulogne Billancourt, France}
\author{Evgeny A. Stepanov}
\affiliation{Radboud University, Institute for Molecules and Materials, NL-6525 AJ Nijmegen, The Netherlands}
\author{Alexander I. Lichtenstein}
\affiliation{Institute of Theoretical Physics, University of Hamburg, 20355 Hamburg, Germany}
\author{Mikhail I. Katsnelson}
\affiliation{Radboud University, Institute for Molecules and Materials, NL-6525 AJ Nijmegen, The Netherlands}

\begin{abstract}
We discuss the calculation of the double occupancy using Dynamical Mean-Field Theory (DMFT) in \emph{finite} dimensions. The double occupancy can be determined from the susceptibility of the auxiliary impurity model or from the lattice susceptibility. 
The former method typically overestimates, whereas the latter underestimates the double occupancy. We illustrate this for the square-lattice Hubbard model. We propose an approach for which both methods lead to identical results by construction and which resolves this ambiguity. This self-consistent dual boson scheme results in a double occupancy that is numerically close to benchmarks available in the literature.
\end{abstract}

\date{\today}

\maketitle
 
 The double occupancy plays an important role in the study of correlated systems.
 It is indicative of the Mott metal-insulator transition and of local moment formation.
 In optical lattice experiments, the double occupancy~\cite{Jordens08,Scarola09,Strohmaier10} gives information about the phase.
 On the theory side, the double occupancy has been used to benchmark approximations~\cite{leblanc15}. It also enters the calculation of total energies and forces in LDA+DMFT studies of strongly correlated materials~\cite{Leonov14}. 
 Given this important role, it is worthwhile to look at how the double occupancy is determined.
 Here, we consider the double occupancy in DMFT and its extension Dual Boson (DB). We compare our results with benchmark results~\cite{leblanc15} available in the literature. 
 
 Over the past two decades, DMFT~\cite{Metzner89,Georges96} has become the dominant approximation for strongly correlated electron systems~\cite{Kotliar06}.
 DMFT solves a self-consistently determined auxiliary single-site problem (the impurity problem) to determine properties of the original lattice problem. Initially, DMFT studies focused on infinite-dimensional systems, where the approximation becomes exact. In this case, the double occupancy is given by the double occupancy of the auxiliary impurity problem~\cite{Rozenberg99}.
Nowadays, DMFT has become an accepted approximation also for finite-dimensional systems. In these calculations, it is often assumed that the double occupancy is equal to that of the impurity problem\footnote{Most papers are not very explicit in how they calculate the double occupancy. 
 An alternative expression~\cite{Gull12} is 
 $d=\int_{\kv\nu} 2T/U~ \Sigma_{\kv\nu} G_{\kv\nu}$, based on the Galitskii-Migdal formula for the total energy~\cite{Galitskii58,DiMarco09}.
 In infinite dimensions and in the DMFT approximation, the self-energy is local, and the formula simplifies to
 $d= \int_\nu  2T/U ~ \Sigma_\nu G^\text{loc}_{\nu}$.
 This yields exactly the same double occupancy as the impurity problem, since the impurity problem is solved exactly.
 }. Here we show that this assumption is potentially problematic. 

To be concrete, we study the Hubbard model as a prototypical example of a strongly correlated system. It is described by the Hamiltonian
 \begin{align}
 H =& -t \sum_{\av{ab}\sigma}c^\dagger_{b\sigma}c^{\phantom{\dagger}}_{a\sigma}+\sum_{a}U n_{a\up} n_{a\dn},
 \end{align}
 where $t$ is the hopping amplitude of an electron with spin $\sigma$ between nearest-neighbor sites $a$ and $b$, and $U$ is the on-site repulsion between electrons. The local repulsion $U$ disfavors doubly-occupied sites and hence decreases the double occupancy $d=\av{n_\up n_\dn}$. 
 
 The double occupancy is a local two-particle correlation function. It can be written as the difference between
the charge and spin susceptibility. In obvious short-hand notation, 
\begin{align}
 d =& \frac{1}{4} \left[\av{nn}-\av{S_z S_z}\right]\notag\\
 =& \frac{1}{4} \left[\Xch_{\text{loc}}-\Xsz_{\text{loc}}+\av{n}\av{n}-\av{S_z}\av{S_z} \right].\label{eq:double}
\end{align}
Here $\Xch_{\text{loc}} = \av{nn} - \av{n}\av{n}$ and $\Xsz_{\text{loc}} = \av{S_z S_z} -\av{S_z}\av{S_z}$ are the equal-time, local correlation functions of the charge density $n=n_\up + n_\dn$ and the magnetization $S_z = n_\up - n_\dn$, respectively~\footnote{To emphasize the analogy between magnetic and charge susceptibilities, we define magnetization without a factor $\frac{1}{2}$.}.
In DMFT, the charge and spin susceptibilities in turn are either approximated by the respective impurity susceptibility, or by the DMFT lattice susceptibilities. The latter are computed by summing ladder diagrams containing the lattice Green's function and a local irreducible vertex~\cite{Brandt89,Georges96}. The computation of the DMFT double occupancy is ambiguous, because the two-particle impurity correlation functions are not identical to the local lattice correlation functions. Recently, we showed this for the compressibility~\cite{vanLoon15}.

\begin{figure}
  \includegraphics{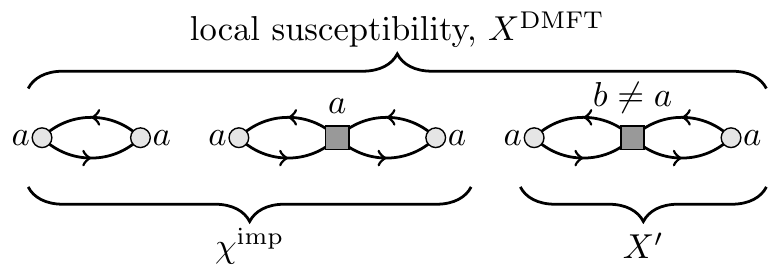}
  \caption{Visual illustration of Eq.~\eqref{eq:xfrompi}: Shown are example diagrams contributing to the local part of the DMFT susceptibility. The impurity susceptibility $\chi^\text{imp}$ contains the local bubble and local vertex corrections built from local propagators. $X'$ contains all the remaining diagrams. These describe processes that start and end at a site $a$ but involve intermediate scattering on different sites $b$ and nonlocal propagators.}
  \label{fig:diagrams}
\end{figure}

To emphasize their difference, we can decompose the susceptibility into the impurity susceptibility $\chiimp$ and a remainder $X'$~\cite{Rubtsov12,Hafermann14-2},
\begin{align}
 X_{\text{DMFT}} = \chiimp + X' \label{eq:xfrompi}.
\end{align}
As illustrated in Fig.~\ref{fig:diagrams}, the impurity susceptibility $\chiimp$ contains two-particle ladder diagrams with only \emph{local} propagators. The local part of the lattice susceptibility $X$ contains ladder diagrams with identical end-points but with nonlocal propagators. These nonlocal vertex contributions are contained in $X'$, for which an explicit expression is available~\cite{Hafermann14-2}.

\begin{figure}
\includegraphics{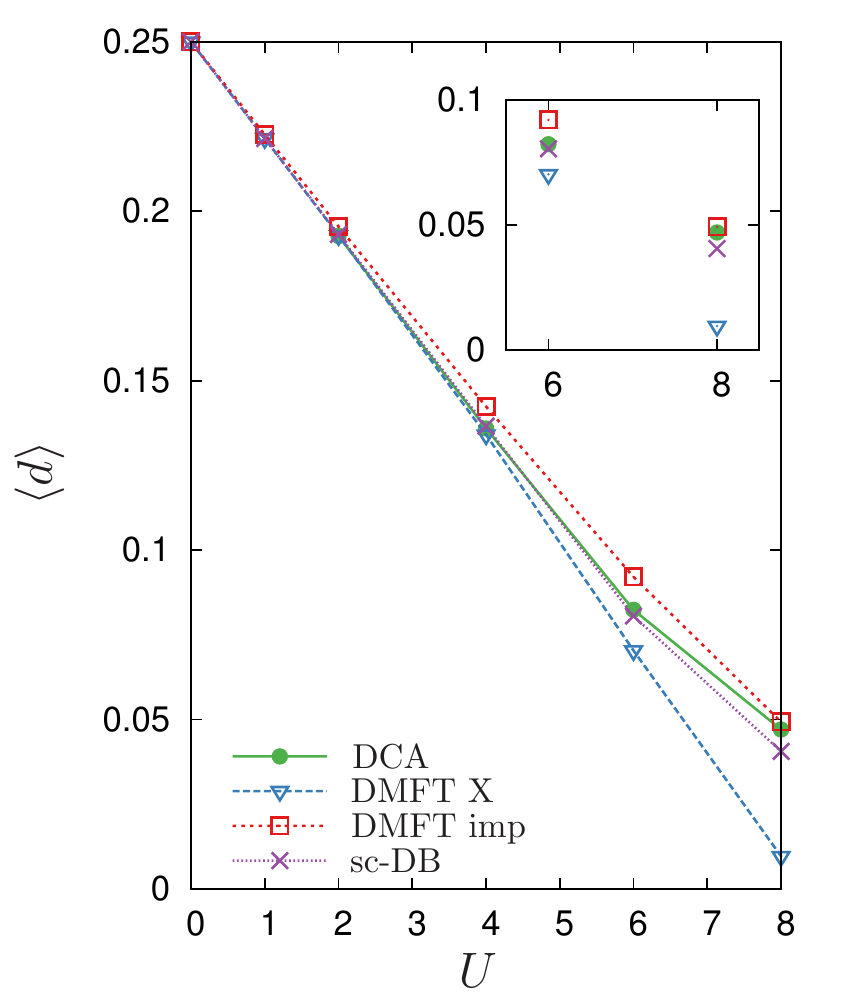}
 \caption{Double occupancy. The lines show the DMFT double occupancy according to Eq.~\eqref{eq:double} (blue triangles) and the impurity double occupancy in DMFT (red squares). The purple crosses show the result of self-consistent dual boson. The double occupancy from DCA is shown~\cite{leblanc15} as a benchmark (green circles).}
 \label{fig:results:d}
\end{figure}

In order to see their effect, we show the double occupancy of the two-dimensional (2d) half-filled square lattice Hubbard model with $t=1$, $\beta=2$ as a function of $U$ in Fig.~\ref{fig:results:d}, determined both from the the DMFT lattice susceptibility \eqref{eq:xfrompi} (blue triangles) and the double occupancy of the auxiliary impurity model (red squares).
For comparison, DCA results extrapolated to infinite lattice size from Ref.~\onlinecite{leblanc15} are shown (green circles). They can be considered the numerically exact solution of this model~\cite{leblanc15}.
The nonlocal vertex corrections in \eqref{eq:xfrompi} include additional correlation effects that tend to reduce the double occupancy, making the system more insulating.
On the other hand, due to its mean-field character, DMFT overestimates the N\'eel temperature~\cite{Rohringer11,Schafer15}.
In particular, this means that the DMFT susceptibility (blue triangles) overestimates $\Xsz$ and thus, according to \eqref{eq:double}, underestimates $d$.
This happens particularly at large $U$, where the tendency towards antiferromagnetism is more pronounced and hence the strongest deviations occur.
In fact, we find that the DMFT susceptibility results in a negative double occupancy at sufficiently large values of $U$, which is clearly unphysical.
The double occupancy of the auxiliary impurity, on the other hand, gets close to the DCA results at large $U$. As $U$ increases, the electrons localize and nonlocal corrections are less important. 

We now turn our attention to the self-consistent DB (sc-DB) approach~\cite{Rubtsov12,vanLoon14-2}. The hallmark of this approach is the self-consistency condition~\cite{Stepanov16},
\begin{align}
X_{\text{loc},\omega}=\chiimp_{\omega}, \label{eq:sc}
\end{align}
which resolves the above ambiguity \emph{by construction}.\footnote{Without the self-consistency condition and in absence of nonlocal interaction, the dual boson approach reduces to DMFT, in particular for the Hubbard model~\cite{vanLoon14-2}.}
This self-consistency condition is achieved by introducing a frequency-dependent interaction $\Lambda_\omega$ to the auxiliary impurity model. 
The effect of this interaction is twofold. 
First of all, the auxiliary impurity problem and the associated $\chiimp$ are different from their DMFT values.
Second, the lattice susceptibility has to be calculated as $X^{-1}=X_\text{DMFT}^{-1}+\Lambda$, where $X_\text{DMFT}$ is the usual DMFT susceptibility obtained from the two-particle ladder, and the equation should be understood in momentum and frequency space. It resembles the RPA expression for the susceptibility, where $X_\text{DMFT}$ plays the role of the bare susceptibility and $-\Lambda$ the role of the interaction.~\footnote{By introducing an additional interaction $\Lambda_\omega$ to the impurity part of the problem, the lattice model is left with an interaction $-\Lambda_\omega$.}
We determine a self-consistent field both for the density and the spin ($S_z$) channel, as they differ in general.

The self-consistent dual boson approach described here shares certain characteristics with the two-particle self-consistent approach~\cite{Vilk94,Vilk97,Tremblay12} (TPSC). Both approaches determine an effective interaction for the spin and charge channel according to a self-consistency condition. Note however, that the original TPSC approach takes the RPA as its starting point, sc-DB starts from DMFT. Strong correlation effects are included in DB from the start. Furthermore, the effective interaction in TPSC is static.
The Moriyaesque $\lambda$ correction~\cite{Katanin09} in D$\Gamma$A~\cite{Toschi07} is also somewhat similar in spirit. It, too, is a correction to the susceptibility that is used to impose self-consistency. Like in TPSC, the Moriyaesque correction is static. Therefore, in both cases, these can only fix certain sum-rules. The effective interaction in sc-DB, on the contrary, may be regarded as a frequency-dependent $\Lambda$-correction, which removes the ambiguity of calculating local susceptibilities.

The purple crosses in Fig.~\ref{fig:results:d} show the result of an sc-DB calculation with self-consistently determined frequency-dependent interactions $\Lambda_\omega$ in the density and spin ($S_z$) channel. 
The results are much closer to the DCA benchmark than both DMFT approaches. This shows that taking into account feedback of collective excitations onto the effective impurity problem can lead to a significant improvement. 

We note that it was difficult to converge our scheme for $U>8$. This coincides with the parameter region where the DMFT lattice susceptibility suggests a negative double occupancy. Note that a negative double occupancy cannot occur in sc-DB with the self-consistency condition \eqref{eq:sc}, since the impurity double occupancy is always positive.
We expect that convergence will be improved by also taking into account the feedback of the $S_x$ and $S_y$ channels~\cite{Otsuki13} onto the impurity. These are especially important at larger $U$ where Heisenberg physics is dominant (see also Ref.~\onlinecite{ayral15} for a discussion on the Ising versus Heisenberg decoupling). 

\begin{figure}
\includegraphics{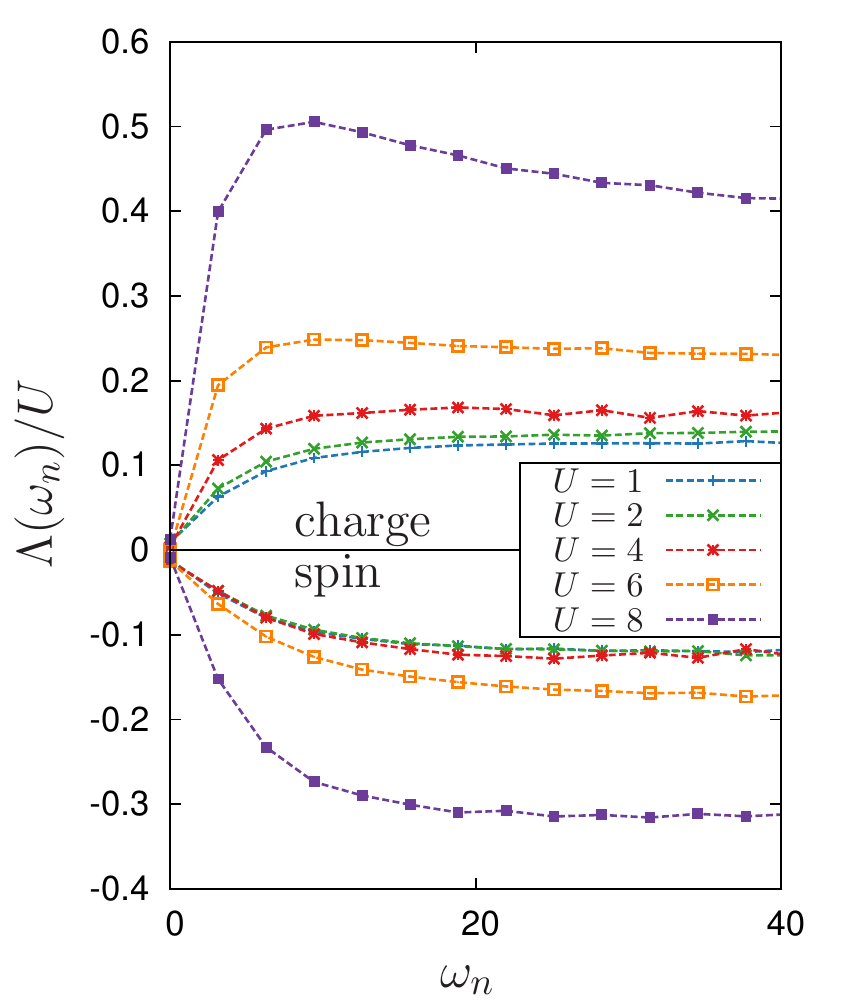}
 \caption{%
 Screening of the impurity interaction in sc-DB. The different values of $U$ correspond to the point in Fig.~\ref{fig:results:d}. The difference between the screened and bare interaction is shown both in the charge and in the spin channel, normalized by the bare interaction.  The retarded interactions appear as $n\Lambda^\text{ch} n + S_z\Lambda^\text{sz}  S_z$ in the impurity action, so both the positive sign in the charge channel and the negative sign in the spin channel suppress the double occupancy.
 }
 \label{fig:results:phiw}
\end{figure}

The effective interactions $\Lambda_\omega$ are crucial to the sc-DB approach. In Fig.~\ref{fig:results:phiw} we show these quantities for the parameters of Fig.~\ref{fig:results:d}. 
At small $U$, the retarded interaction is proportional to $U$~\cite{Stepanov16}. Fig.~\ref{fig:results:phiw} shows that this linear scaling roughly holds until $U\approx 4$ in the spin channel, whereas nonlinear behavior is already visible at $U\approx 2$ in the charge channel.
The retarded interaction goes to a nonzero constant at high frequencies, which in general is different for the charge and spin channels. 
This is somewhat reminiscent of TPSC, where the interaction is a constant (independent of frequency) in both channels, with a different constant for the charge and spin channels. In sc-DB, the renormalization of the effective interaction is largest in the charge channel. This behavior is also observed in TPSC~\cite{Vilk97}. The Hubbard repulsion strongly suppresses charge fluctuations.
At $U=8$, the charge channel develops a small maximum at intermediate frequency. This occurs at the typical energy scale of charge fluctuations, i.e., $\omega_n \approx U$, however the precise physical origin of this maximum is currently unknown.

\begin{figure}[]
\includegraphics{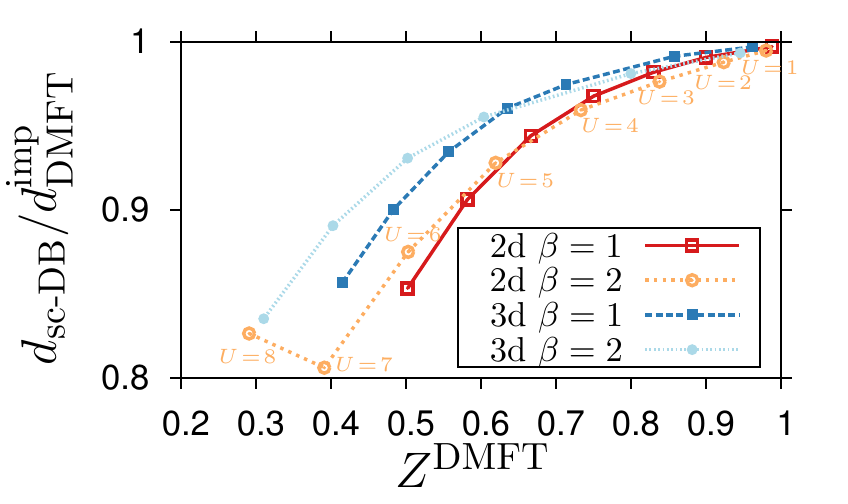}
 \caption{%
Renormalization of the impurity double occupancy by nonlocal processes in two and three dimensions. Shown is the ratio of the sc-DB and DMFT impurity susceptibilities as a function of the quasiparticle weight $Z$. The results in 2d at $\beta=2$ correspond to Fig.~\ref{fig:results:d}. In 2d, the renormalization effect is stronger than in 3d.
 }
 \label{fig:results:dimensions}
\end{figure}

The dimensionality of the model plays an important role in DMFT-based studies~\cite{Metzner89,Georges96}.
To illustrate the effect of the dimensionality, we have performed similar calculations in a 3d simple cubic lattice. The results are shown  in Fig.~\ref{fig:results:dimensions}. 
A direct comparison is difficult by the change in energy scales that occurs  when changing the dimension.
To overcome this, we use the $Z$-factor, $Z=(1 - d\text{Re}\Sigma_{\omega}/d\omega)^{-1}$, obtained in DMFT, to indicate the importance of interaction effects on the one-particle level.
The renormalization of the double occupancy by nonlocal processes, $d_{\text{sc-DB}}/d^{\text{imp}}_{\text{DMFT}}$, is shown as a function of $Z$.
The figure clearly shows that the sc-DB susceptibility contains important nonlocal corrections.
Although it is clear that other factors than dimensionality also play a role, Fig.~\ref{fig:results:dimensions} also suggests that the nonlocal renormalization of the double occupancy is less important in the 3d system. 
This result matches our physical expectation that nonlocal correlation effects are stronger in lower dimensions. 
In fact, vertex corrections to the susceptibility vanish in infinite dimension~\cite{Khurana90,Georges96,Stepanov16}, $X'_{\text{local}}=0$, and $d=d^\text{imp}$ in DMFT, so no ambiguity occurs and there is no nonlocal renormalization of the double occupancy.

Fig.~\ref{fig:results:dimensions} also shows that the nonlocal renormalization is small in weakly interacting systems, with $Z\approx 1$. According to Eq. \ref{eq:xfrompi}, nonlocal vertex corrections are responsible for this nonlocal renormalization, and the vertex is small at small $U$. On the other hand, at very large $U$, the electrons would be very localized and as a result, \emph{nonlocal} vertex corrections are again small. The uptick between $U=7$ to $U=8$ in Fig.~\ref{fig:results:dimensions} is a first sign of this. This behavior is similar to the eigenvalue analysis in dual fermion~\cite{Hafermann09}. The corrections to the local auxiliary impurity are largest at intermediate $U$.

\begin{figure}[]
\includegraphics{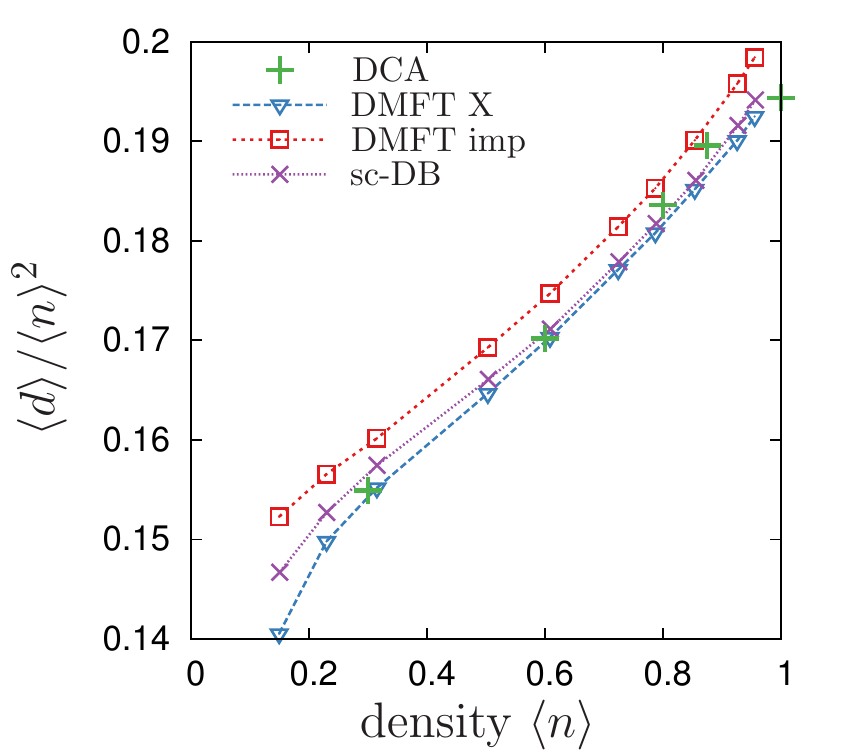}
 \caption{%
Double occupancy away from half-filling at $U=2$, $\beta=8$, in the same color scheme as Fig.~\ref{fig:results:d}. The double occupancy is normalized by $\av{n}^2$ for clarity.
 }
 \label{fig:results:filling}
\end{figure}

In Fig.~\ref{fig:results:filling}, we show the double occupancy similar to Fig.~\ref{fig:results:d}, away from half-filling, at constant $U=2$ and lower temperature $\beta=8$ as a function of $\av{n}$. For clarity, we show the double occupancy normalized by the density squared. For a noninteracting system, $d=\av{n}^2/4$, so the deviation from $1/4$ shows the interaction effects. At this lowered temperature, the half-filled system has strong antiferromagnetic fluctuations that inhibit sc-DB calculations. As in Fig.~\ref{fig:results:d}, both the DCA and sc-DB results lie between the results obtained from the DMFT lattice susceptibility and from the auxiliary impurity model. Again the impurity susceptibility over- and the lattice susceptibility underestimates the double occupancy. The largest inconsistency between the two occurs actually towards lower filling, similar to what happens for the compressibility~\cite{vanLoon15}.
While the DCA benchmarks~\cite{leblanc15} only go down to $\av{n}=0.3$, these results suggest that the region at lower fillings still contains interesting physics.

In conclusion, we have proposed a scheme which resolves the ambiguity in the calculation of the double occupancy inherent to DMFT in finite dimensions.
DMFT is not self-consistent on the two-particle level, and as a result, the single- and two-particle quantities are not compatible, as exemplified here by the difference between the double occupancy from the Galitskii-Migdal formula and from the susceptibility. 
This ambiguity extends to the determination of total energies and forces~\cite{Leonov14} from DMFT.

The self-consistent dual boson approach gives results which are in good agreement with benchmarks in the literature.
We have further shown that the DMFT impurity double occupancy typically overestimates the double occupancy, while the occupancy determined from the DMFT lattice susceptibility underestimates it significantly. For large interaction, this can even lead to unphysical negative results. We have seen that in two-dimensional systems and at a moderate $Z$-factor of $0.5$, the results may differ by 10\%-20\%. The double occupancies determined from the DMFT lattice and impurity susceptibilities differ strongly also away from half filling. The discrepancy may have a significant effect on total energy and force calculations and hence the determination of equilibrium positions of the atoms in realistic LDA+DMFT calculations. This should be kept in mind when interpreting the results. In order to avoid unphysical results we recommend to approximate the double occupancy by the impurity double occupancy in LDA+DMFT. The accuracy of this approach however remains to be determined.

\acknowledgments

The authors thank Lewin Boehnke, Silke Biermann, Philipp Werner, Junya Otsuki, Tim Wehling, Alessandro Toschi and Igor di Marco for useful discussion.
The authors thank James LeBlanc for help with the DCA benchmarks.
E.G.C.P.v.L., E.A.S and M.I.K. acknowledge support from ERC Advanced Grant 338957 FEMTO/NANO. F.K. and A.L. are supported by the DFG-FOR1346 program.
The auxiliary impurity model was solved using a modified version of the open source CT-HYB solver~\cite{Hafermann13,Hafermann14} based on the ALPS libraries~\cite{ALPS2}.

\bibliography{main}

\begin{thebibliography}{36}%
\makeatletter
\providecommand \@ifxundefined [1]{%
 \@ifx{#1\undefined}
}%
\providecommand \@ifnum [1]{%
 \ifnum #1\expandafter \@firstoftwo
 \else \expandafter \@secondoftwo
 \fi
}%
\providecommand \@ifx [1]{%
 \ifx #1\expandafter \@firstoftwo
 \else \expandafter \@secondoftwo
 \fi
}%
\providecommand \natexlab [1]{#1}%
\providecommand \enquote  [1]{``#1''}%
\providecommand \bibnamefont  [1]{#1}%
\providecommand \bibfnamefont [1]{#1}%
\providecommand \citenamefont [1]{#1}%
\providecommand \href@noop [0]{\@secondoftwo}%
\providecommand \href [0]{\begingroup \@sanitize@url \@href}%
\providecommand \@href[1]{\@@startlink{#1}\@@href}%
\providecommand \@@href[1]{\endgroup#1\@@endlink}%
\providecommand \@sanitize@url [0]{\catcode `\\12\catcode `\$12\catcode
  `\&12\catcode `\#12\catcode `\^12\catcode `\_12\catcode `\%12\relax}%
\providecommand \@@startlink[1]{}%
\providecommand \@@endlink[0]{}%
\providecommand \url  [0]{\begingroup\@sanitize@url \@url }%
\providecommand \@url [1]{\endgroup\@href {#1}{\urlprefix }}%
\providecommand \urlprefix  [0]{URL }%
\providecommand \Eprint [0]{\href }%
\providecommand \doibase [0]{http://dx.doi.org/}%
\providecommand \selectlanguage [0]{\@gobble}%
\providecommand \bibinfo  [0]{\@secondoftwo}%
\providecommand \bibfield  [0]{\@secondoftwo}%
\providecommand \translation [1]{[#1]}%
\providecommand \BibitemOpen [0]{}%
\providecommand \bibitemStop [0]{}%
\providecommand \bibitemNoStop [0]{.\EOS\space}%
\providecommand \EOS [0]{\spacefactor3000\relax}%
\providecommand \BibitemShut  [1]{\csname bibitem#1\endcsname}%
\let\auto@bib@innerbib\@empty
\bibitem [{\citenamefont {{Robert J\"ordens}}\ \emph
  {et~al.}(2008)\citenamefont {{Robert J\"ordens}}, \citenamefont {{Niels
  Strohmaier}}, \citenamefont {{Kenneth G\"unter }}, \citenamefont {{Henning
  Moritz}},\ and\ \citenamefont {{Tilman Esslinger}}}]{Jordens08}%
  \BibitemOpen
  \bibfield  {author} {\bibinfo {author} {\bibnamefont {{Robert J\"ordens}}},
  \bibinfo {author} {\bibnamefont {{Niels Strohmaier}}}, \bibinfo {author}
  {\bibnamefont {{Kenneth G\"unter }}}, \bibinfo {author} {\bibnamefont
  {{Henning Moritz}}}, \ and\ \bibinfo {author} {\bibnamefont {{Tilman
  Esslinger}}},\ }\href {\doibase http://dx.doi.org/10.1038/nature07244}
  {\bibfield  {journal} {\bibinfo  {journal} {Nature}\ }\textbf {\bibinfo
  {volume} {455}},\ \bibinfo {pages} {204} (\bibinfo {year} {2008})},\ \bibinfo
  {note} {10.1038/nature07244}\BibitemShut {NoStop}%
\bibitem [{\citenamefont {Scarola}\ \emph {et~al.}(2009)\citenamefont
  {Scarola}, \citenamefont {Pollet}, \citenamefont {Oitmaa},\ and\
  \citenamefont {Troyer}}]{Scarola09}%
  \BibitemOpen
  \bibfield  {author} {\bibinfo {author} {\bibfnamefont {V.~W.}\ \bibnamefont
  {Scarola}}, \bibinfo {author} {\bibfnamefont {L.}~\bibnamefont {Pollet}},
  \bibinfo {author} {\bibfnamefont {J.}~\bibnamefont {Oitmaa}}, \ and\ \bibinfo
  {author} {\bibfnamefont {M.}~\bibnamefont {Troyer}},\ }\href {\doibase
  10.1103/PhysRevLett.102.135302} {\bibfield  {journal} {\bibinfo  {journal}
  {Phys. Rev. Lett.}\ }\textbf {\bibinfo {volume} {102}},\ \bibinfo {pages}
  {135302} (\bibinfo {year} {2009})}\BibitemShut {NoStop}%
\bibitem [{\citenamefont {Strohmaier}\ \emph {et~al.}(2010)\citenamefont
  {Strohmaier}, \citenamefont {Greif}, \citenamefont {J\"ordens}, \citenamefont
  {Tarruell}, \citenamefont {Moritz}, \citenamefont {Esslinger}, \citenamefont
  {Sensarma}, \citenamefont {Pekker}, \citenamefont {Altman},\ and\
  \citenamefont {Demler}}]{Strohmaier10}%
  \BibitemOpen
  \bibfield  {author} {\bibinfo {author} {\bibfnamefont {N.}~\bibnamefont
  {Strohmaier}}, \bibinfo {author} {\bibfnamefont {D.}~\bibnamefont {Greif}},
  \bibinfo {author} {\bibfnamefont {R.}~\bibnamefont {J\"ordens}}, \bibinfo
  {author} {\bibfnamefont {L.}~\bibnamefont {Tarruell}}, \bibinfo {author}
  {\bibfnamefont {H.}~\bibnamefont {Moritz}}, \bibinfo {author} {\bibfnamefont
  {T.}~\bibnamefont {Esslinger}}, \bibinfo {author} {\bibfnamefont
  {R.}~\bibnamefont {Sensarma}}, \bibinfo {author} {\bibfnamefont
  {D.}~\bibnamefont {Pekker}}, \bibinfo {author} {\bibfnamefont
  {E.}~\bibnamefont {Altman}}, \ and\ \bibinfo {author} {\bibfnamefont
  {E.}~\bibnamefont {Demler}},\ }\href {\doibase
  10.1103/PhysRevLett.104.080401} {\bibfield  {journal} {\bibinfo  {journal}
  {Phys. Rev. Lett.}\ }\textbf {\bibinfo {volume} {104}},\ \bibinfo {pages}
  {080401} (\bibinfo {year} {2010})}\BibitemShut {NoStop}%
\bibitem [{\citenamefont {LeBlanc}\ \emph {et~al.}(2015)\citenamefont
  {LeBlanc}, \citenamefont {Antipov}, \citenamefont {Becca}, \citenamefont
  {Bulik}, \citenamefont {Chan}, \citenamefont {Chung}, \citenamefont {Deng},
  \citenamefont {Ferrero}, \citenamefont {Henderson}, \citenamefont
  {Jim\'enez-Hoyos}, \citenamefont {Kozik}, \citenamefont {Liu}, \citenamefont
  {Millis}, \citenamefont {Prokof'ev}, \citenamefont {Qin}, \citenamefont
  {Scuseria}, \citenamefont {Shi}, \citenamefont {Svistunov}, \citenamefont
  {Tocchio}, \citenamefont {Tupitsyn}, \citenamefont {White}, \citenamefont
  {Zhang}, \citenamefont {Zheng}, \citenamefont {Zhu},\ and\ \citenamefont
  {Gull}}]{leblanc15}%
  \BibitemOpen
  \bibfield  {author} {\bibinfo {author} {\bibfnamefont {J.~P.~F.}\
  \bibnamefont {LeBlanc}}, \bibinfo {author} {\bibfnamefont {A.~E.}\
  \bibnamefont {Antipov}}, \bibinfo {author} {\bibfnamefont {F.}~\bibnamefont
  {Becca}}, \bibinfo {author} {\bibfnamefont {I.~W.}\ \bibnamefont {Bulik}},
  \bibinfo {author} {\bibfnamefont {G.~K.-L.}\ \bibnamefont {Chan}}, \bibinfo
  {author} {\bibfnamefont {C.-M.}\ \bibnamefont {Chung}}, \bibinfo {author}
  {\bibfnamefont {Y.}~\bibnamefont {Deng}}, \bibinfo {author} {\bibfnamefont
  {M.}~\bibnamefont {Ferrero}}, \bibinfo {author} {\bibfnamefont {T.~M.}\
  \bibnamefont {Henderson}}, \bibinfo {author} {\bibfnamefont {C.~A.}\
  \bibnamefont {Jim\'enez-Hoyos}}, \bibinfo {author} {\bibfnamefont
  {E.}~\bibnamefont {Kozik}}, \bibinfo {author} {\bibfnamefont {X.-W.}\
  \bibnamefont {Liu}}, \bibinfo {author} {\bibfnamefont {A.~J.}\ \bibnamefont
  {Millis}}, \bibinfo {author} {\bibfnamefont {N.~V.}\ \bibnamefont
  {Prokof'ev}}, \bibinfo {author} {\bibfnamefont {M.}~\bibnamefont {Qin}},
  \bibinfo {author} {\bibfnamefont {G.~E.}\ \bibnamefont {Scuseria}}, \bibinfo
  {author} {\bibfnamefont {H.}~\bibnamefont {Shi}}, \bibinfo {author}
  {\bibfnamefont {B.~V.}\ \bibnamefont {Svistunov}}, \bibinfo {author}
  {\bibfnamefont {L.~F.}\ \bibnamefont {Tocchio}}, \bibinfo {author}
  {\bibfnamefont {I.~S.}\ \bibnamefont {Tupitsyn}}, \bibinfo {author}
  {\bibfnamefont {S.~R.}\ \bibnamefont {White}}, \bibinfo {author}
  {\bibfnamefont {S.}~\bibnamefont {Zhang}}, \bibinfo {author} {\bibfnamefont
  {B.-X.}\ \bibnamefont {Zheng}}, \bibinfo {author} {\bibfnamefont
  {Z.}~\bibnamefont {Zhu}}, \ and\ \bibinfo {author} {\bibfnamefont
  {E.}~\bibnamefont {Gull}} (\bibinfo {collaboration} {Simons Collaboration on
  the Many-Electron Problem}),\ }\href {\doibase 10.1103/PhysRevX.5.041041}
  {\bibfield  {journal} {\bibinfo  {journal} {Phys. Rev. X}\ }\textbf {\bibinfo
  {volume} {5}},\ \bibinfo {pages} {041041} (\bibinfo {year}
  {2015})}\BibitemShut {NoStop}%
\bibitem [{\citenamefont {Leonov}\ \emph {et~al.}(2014)\citenamefont {Leonov},
  \citenamefont {Anisimov},\ and\ \citenamefont {Vollhardt}}]{Leonov14}%
  \BibitemOpen
  \bibfield  {author} {\bibinfo {author} {\bibfnamefont {I.}~\bibnamefont
  {Leonov}}, \bibinfo {author} {\bibfnamefont {V.~I.}\ \bibnamefont
  {Anisimov}}, \ and\ \bibinfo {author} {\bibfnamefont {D.}~\bibnamefont
  {Vollhardt}},\ }\href {\doibase 10.1103/PhysRevLett.112.146401} {\bibfield
  {journal} {\bibinfo  {journal} {Phys. Rev. Lett.}\ }\textbf {\bibinfo
  {volume} {112}},\ \bibinfo {pages} {146401} (\bibinfo {year}
  {2014})}\BibitemShut {NoStop}%
\bibitem [{\citenamefont {Metzner}\ and\ \citenamefont
  {Vollhardt}(1989)}]{Metzner89}%
  \BibitemOpen
  \bibfield  {author} {\bibinfo {author} {\bibfnamefont {W.}~\bibnamefont
  {Metzner}}\ and\ \bibinfo {author} {\bibfnamefont {D.}~\bibnamefont
  {Vollhardt}},\ }\href {\doibase 10.1103/PhysRevLett.62.324} {\bibfield
  {journal} {\bibinfo  {journal} {Phys. Rev. Lett.}\ }\textbf {\bibinfo
  {volume} {62}},\ \bibinfo {pages} {324} (\bibinfo {year} {1989})}\BibitemShut
  {NoStop}%
\bibitem [{\citenamefont {Georges}\ \emph {et~al.}(1996)\citenamefont
  {Georges}, \citenamefont {Kotliar}, \citenamefont {Krauth},\ and\
  \citenamefont {Rozenberg}}]{Georges96}%
  \BibitemOpen
  \bibfield  {author} {\bibinfo {author} {\bibfnamefont {A.}~\bibnamefont
  {Georges}}, \bibinfo {author} {\bibfnamefont {G.}~\bibnamefont {Kotliar}},
  \bibinfo {author} {\bibfnamefont {W.}~\bibnamefont {Krauth}}, \ and\ \bibinfo
  {author} {\bibfnamefont {M.~J.}\ \bibnamefont {Rozenberg}},\ }\href {\doibase
  10.1103/RevModPhys.68.13} {\bibfield  {journal} {\bibinfo  {journal} {Rev.
  Mod. Phys.}\ }\textbf {\bibinfo {volume} {68}},\ \bibinfo {pages} {13}
  (\bibinfo {year} {1996})}\BibitemShut {NoStop}%
\bibitem [{\citenamefont {Kotliar}\ \emph {et~al.}(2006)\citenamefont
  {Kotliar}, \citenamefont {Savrasov}, \citenamefont {Haule}, \citenamefont
  {Oudovenko}, \citenamefont {Parcollet},\ and\ \citenamefont
  {Marianetti}}]{Kotliar06}%
  \BibitemOpen
  \bibfield  {author} {\bibinfo {author} {\bibfnamefont {G.}~\bibnamefont
  {Kotliar}}, \bibinfo {author} {\bibfnamefont {S.~Y.}\ \bibnamefont
  {Savrasov}}, \bibinfo {author} {\bibfnamefont {K.}~\bibnamefont {Haule}},
  \bibinfo {author} {\bibfnamefont {V.~S.}\ \bibnamefont {Oudovenko}}, \bibinfo
  {author} {\bibfnamefont {O.}~\bibnamefont {Parcollet}}, \ and\ \bibinfo
  {author} {\bibfnamefont {C.~A.}\ \bibnamefont {Marianetti}},\ }\href
  {\doibase 10.1103/RevModPhys.78.865} {\bibfield  {journal} {\bibinfo
  {journal} {Rev. Mod. Phys.}\ }\textbf {\bibinfo {volume} {78}},\ \bibinfo
  {pages} {865} (\bibinfo {year} {2006})}\BibitemShut {NoStop}%
\bibitem [{\citenamefont {Rozenberg}\ \emph {et~al.}(1999)\citenamefont
  {Rozenberg}, \citenamefont {Chitra},\ and\ \citenamefont
  {Kotliar}}]{Rozenberg99}%
  \BibitemOpen
  \bibfield  {author} {\bibinfo {author} {\bibfnamefont {M.~J.}\ \bibnamefont
  {Rozenberg}}, \bibinfo {author} {\bibfnamefont {R.}~\bibnamefont {Chitra}}, \
  and\ \bibinfo {author} {\bibfnamefont {G.}~\bibnamefont {Kotliar}},\ }\href
  {\doibase 10.1103/PhysRevLett.83.3498} {\bibfield  {journal} {\bibinfo
  {journal} {Phys. Rev. Lett.}\ }\textbf {\bibinfo {volume} {83}},\ \bibinfo
  {pages} {3498} (\bibinfo {year} {1999})}\BibitemShut {NoStop}%
\bibitem [{Note1()}]{Note1}%
  \BibitemOpen
  \bibinfo {note} {Most papers are not very explicit in how they calculate the
  double occupancy. An alternative expression~\cite {Gull12} is $d=\DOTSI
  \intop \ilimits@ _{\protect \ensuremath {\protect \mathbf {k}}\nu } 2T/U~
  \Sigma _{\protect \ensuremath {\protect \mathbf {k}}\nu } G_{\protect
  \ensuremath {\protect \mathbf {k}}\nu }$, based on the Galitskii-Migdal
  formula for the total energy~\cite {Galitskii58,DiMarco09}. In infinite
  dimensions and in the DMFT approximation, the self-energy is local, and the
  formula simplifies to $d= \DOTSI \intop \ilimits@ _\nu 2T/U ~ \Sigma _\nu
  G^\protect \text {loc}_{\nu }$. This yields exactly the same double occupancy
  as the impurity problem, since the impurity problem is solved
  exactly.}\BibitemShut {Stop}%
\bibitem [{Note2()}]{Note2}%
  \BibitemOpen
  \bibinfo {note} {To emphasize the analogy between magnetic and charge
  susceptibilities, we define magnetization without a factor $\protect \frac
  {1}{2}$.}\BibitemShut {Stop}%
\bibitem [{\citenamefont {Brandt}\ and\ \citenamefont
  {Mielsch}(1989)}]{Brandt89}%
  \BibitemOpen
  \bibfield  {author} {\bibinfo {author} {\bibfnamefont {U.}~\bibnamefont
  {Brandt}}\ and\ \bibinfo {author} {\bibfnamefont {C.}~\bibnamefont
  {Mielsch}},\ }\href {\doibase 10.1007/BF01321824} {\bibfield  {journal}
  {\bibinfo  {journal} {Zeitschrift für Physik B Condensed Matter}\ }\textbf
  {\bibinfo {volume} {75}},\ \bibinfo {pages} {365} (\bibinfo {year}
  {1989})}\BibitemShut {NoStop}%
\bibitem [{\citenamefont {van Loon}\ \emph {et~al.}(2015)\citenamefont {van
  Loon}, \citenamefont {Hafermann}, \citenamefont {Lichtenstein},\ and\
  \citenamefont {Katsnelson}}]{vanLoon15}%
  \BibitemOpen
  \bibfield  {author} {\bibinfo {author} {\bibfnamefont {E.~G. C.~P.}\
  \bibnamefont {van Loon}}, \bibinfo {author} {\bibfnamefont {H.}~\bibnamefont
  {Hafermann}}, \bibinfo {author} {\bibfnamefont {A.~I.}\ \bibnamefont
  {Lichtenstein}}, \ and\ \bibinfo {author} {\bibfnamefont {M.~I.}\
  \bibnamefont {Katsnelson}},\ }\href {\doibase 10.1103/PhysRevB.92.085106}
  {\bibfield  {journal} {\bibinfo  {journal} {Phys. Rev. B}\ }\textbf {\bibinfo
  {volume} {92}},\ \bibinfo {pages} {085106} (\bibinfo {year}
  {2015})}\BibitemShut {NoStop}%
\bibitem [{\citenamefont {Rubtsov}\ \emph {et~al.}(2012)\citenamefont
  {Rubtsov}, \citenamefont {Katsnelson},\ and\ \citenamefont
  {Lichtenstein}}]{Rubtsov12}%
  \BibitemOpen
  \bibfield  {author} {\bibinfo {author} {\bibfnamefont {A.~N.}\ \bibnamefont
  {Rubtsov}}, \bibinfo {author} {\bibfnamefont {M.~I.}\ \bibnamefont
  {Katsnelson}}, \ and\ \bibinfo {author} {\bibfnamefont {A.~I.}\ \bibnamefont
  {Lichtenstein}},\ }\href {\doibase 10.1016/j.aop.2012.01.002} {\bibfield
  {journal} {\bibinfo  {journal} {Annals of Physics}\ }\textbf {\bibinfo
  {volume} {327}},\ \bibinfo {pages} {1320} (\bibinfo {year}
  {2012})}\BibitemShut {NoStop}%
\bibitem [{\citenamefont {Hafermann}\ \emph {et~al.}(2014)\citenamefont
  {Hafermann}, \citenamefont {van Loon}, \citenamefont {Katsnelson},
  \citenamefont {Lichtenstein},\ and\ \citenamefont
  {Parcollet}}]{Hafermann14-2}%
  \BibitemOpen
  \bibfield  {author} {\bibinfo {author} {\bibfnamefont {H.}~\bibnamefont
  {Hafermann}}, \bibinfo {author} {\bibfnamefont {E.~G. C.~P.}\ \bibnamefont
  {van Loon}}, \bibinfo {author} {\bibfnamefont {M.~I.}\ \bibnamefont
  {Katsnelson}}, \bibinfo {author} {\bibfnamefont {A.~I.}\ \bibnamefont
  {Lichtenstein}}, \ and\ \bibinfo {author} {\bibfnamefont {O.}~\bibnamefont
  {Parcollet}},\ }\href {\doibase 10.1103/PhysRevB.90.235105} {\bibfield
  {journal} {\bibinfo  {journal} {Phys. Rev. B}\ }\textbf {\bibinfo {volume}
  {90}},\ \bibinfo {pages} {235105} (\bibinfo {year} {2014})}\BibitemShut
  {NoStop}%
\bibitem [{\citenamefont {Rohringer}\ \emph {et~al.}(2011)\citenamefont
  {Rohringer}, \citenamefont {Toschi}, \citenamefont {Katanin},\ and\
  \citenamefont {Held}}]{Rohringer11}%
  \BibitemOpen
  \bibfield  {author} {\bibinfo {author} {\bibfnamefont {G.}~\bibnamefont
  {Rohringer}}, \bibinfo {author} {\bibfnamefont {A.}~\bibnamefont {Toschi}},
  \bibinfo {author} {\bibfnamefont {A.}~\bibnamefont {Katanin}}, \ and\
  \bibinfo {author} {\bibfnamefont {K.}~\bibnamefont {Held}},\ }\href {\doibase
  10.1103/PhysRevLett.107.256402} {\bibfield  {journal} {\bibinfo  {journal}
  {Phys. Rev. Lett.}\ }\textbf {\bibinfo {volume} {107}},\ \bibinfo {pages}
  {256402} (\bibinfo {year} {2011})}\BibitemShut {NoStop}%
\bibitem [{\citenamefont {Sch\"afer}\ \emph {et~al.}(2015)\citenamefont
  {Sch\"afer}, \citenamefont {Geles}, \citenamefont {Rost}, \citenamefont
  {Rohringer}, \citenamefont {Arrigoni}, \citenamefont {Held}, \citenamefont
  {Bl\"umer}, \citenamefont {Aichhorn},\ and\ \citenamefont
  {Toschi}}]{Schafer15}%
  \BibitemOpen
  \bibfield  {author} {\bibinfo {author} {\bibfnamefont {T.}~\bibnamefont
  {Sch\"afer}}, \bibinfo {author} {\bibfnamefont {F.}~\bibnamefont {Geles}},
  \bibinfo {author} {\bibfnamefont {D.}~\bibnamefont {Rost}}, \bibinfo {author}
  {\bibfnamefont {G.}~\bibnamefont {Rohringer}}, \bibinfo {author}
  {\bibfnamefont {E.}~\bibnamefont {Arrigoni}}, \bibinfo {author}
  {\bibfnamefont {K.}~\bibnamefont {Held}}, \bibinfo {author} {\bibfnamefont
  {N.}~\bibnamefont {Bl\"umer}}, \bibinfo {author} {\bibfnamefont
  {M.}~\bibnamefont {Aichhorn}}, \ and\ \bibinfo {author} {\bibfnamefont
  {A.}~\bibnamefont {Toschi}},\ }\href {\doibase 10.1103/PhysRevB.91.125109}
  {\bibfield  {journal} {\bibinfo  {journal} {Phys. Rev. B}\ }\textbf {\bibinfo
  {volume} {91}},\ \bibinfo {pages} {125109} (\bibinfo {year}
  {2015})}\BibitemShut {NoStop}%
\bibitem [{\citenamefont {van Loon}\ \emph {et~al.}(2014)\citenamefont {van
  Loon}, \citenamefont {Lichtenstein}, \citenamefont {Katsnelson},
  \citenamefont {Parcollet},\ and\ \citenamefont {Hafermann}}]{vanLoon14-2}%
  \BibitemOpen
  \bibfield  {author} {\bibinfo {author} {\bibfnamefont {E.~G. C.~P.}\
  \bibnamefont {van Loon}}, \bibinfo {author} {\bibfnamefont {A.~I.}\
  \bibnamefont {Lichtenstein}}, \bibinfo {author} {\bibfnamefont {M.~I.}\
  \bibnamefont {Katsnelson}}, \bibinfo {author} {\bibfnamefont
  {O.}~\bibnamefont {Parcollet}}, \ and\ \bibinfo {author} {\bibfnamefont
  {H.}~\bibnamefont {Hafermann}},\ }\href {\doibase 10.1103/PhysRevB.90.235135}
  {\bibfield  {journal} {\bibinfo  {journal} {Phys. Rev. B}\ }\textbf {\bibinfo
  {volume} {90}},\ \bibinfo {pages} {235135} (\bibinfo {year}
  {2014})}\BibitemShut {NoStop}%
\bibitem [{\citenamefont {Stepanov}\ \emph {et~al.}(2016)\citenamefont
  {Stepanov}, \citenamefont {van Loon}, \citenamefont {Katanin}, \citenamefont
  {Lichtenstein}, \citenamefont {Katsnelson},\ and\ \citenamefont
  {Rubtsov}}]{Stepanov16}%
  \BibitemOpen
  \bibfield  {author} {\bibinfo {author} {\bibfnamefont {E.~A.}\ \bibnamefont
  {Stepanov}}, \bibinfo {author} {\bibfnamefont {E.~G. C.~P.}\ \bibnamefont
  {van Loon}}, \bibinfo {author} {\bibfnamefont {A.~A.}\ \bibnamefont
  {Katanin}}, \bibinfo {author} {\bibfnamefont {A.~I.}\ \bibnamefont
  {Lichtenstein}}, \bibinfo {author} {\bibfnamefont {M.~I.}\ \bibnamefont
  {Katsnelson}}, \ and\ \bibinfo {author} {\bibfnamefont {A.~N.}\ \bibnamefont
  {Rubtsov}},\ }\href {\doibase 10.1103/PhysRevB.93.045107} {\bibfield
  {journal} {\bibinfo  {journal} {Phys. Rev. B}\ }\textbf {\bibinfo {volume}
  {93}},\ \bibinfo {pages} {045107} (\bibinfo {year} {2016})}\BibitemShut
  {NoStop}%
\bibitem [{Note3()}]{Note3}%
  \BibitemOpen
  \bibinfo {note} {Without the self-consistency condition and in absence of
  nonlocal interaction, the dual boson approach reduces to DMFT, in particular
  for the Hubbard model~\cite {vanLoon14-2}.}\BibitemShut {Stop}%
\bibitem [{Note4()}]{Note4}%
  \BibitemOpen
  \bibinfo {note} {By introducing an additional interaction $\Lambda _\omega $
  to the impurity part of the problem, the lattice model is left with an
  interaction $-\Lambda _\omega $.}\BibitemShut {Stop}%
\bibitem [{\citenamefont {Vilk}\ \emph {et~al.}(1994)\citenamefont {Vilk},
  \citenamefont {Chen},\ and\ \citenamefont {Tremblay}}]{Vilk94}%
  \BibitemOpen
  \bibfield  {author} {\bibinfo {author} {\bibfnamefont {Y.~M.}\ \bibnamefont
  {Vilk}}, \bibinfo {author} {\bibfnamefont {L.}~\bibnamefont {Chen}}, \ and\
  \bibinfo {author} {\bibfnamefont {A.-M.~S.}\ \bibnamefont {Tremblay}},\
  }\href {\doibase 10.1103/PhysRevB.49.13267} {\bibfield  {journal} {\bibinfo
  {journal} {Phys. Rev. B}\ }\textbf {\bibinfo {volume} {49}},\ \bibinfo
  {pages} {13267} (\bibinfo {year} {1994})}\BibitemShut {NoStop}%
\bibitem [{\citenamefont {{Y.M. Vilk}}\ and\ \citenamefont {{A.-M.S.
  Tremblay}}(1997)}]{Vilk97}%
  \BibitemOpen
  \bibfield  {author} {\bibinfo {author} {\bibnamefont {{Y.M. Vilk}}}\ and\
  \bibinfo {author} {\bibnamefont {{A.-M.S. Tremblay}}},\ }\href {\doibase
  10.1051/jp1:1997135} {\bibfield  {journal} {\bibinfo  {journal} {J. Phys. I
  France}\ }\textbf {\bibinfo {volume} {7}},\ \bibinfo {pages} {1309} (\bibinfo
  {year} {1997})}\BibitemShut {NoStop}%
\bibitem [{\citenamefont {Tremblay}(2012)}]{Tremblay12}%
  \BibitemOpen
  \bibfield  {author} {\bibinfo {author} {\bibfnamefont {A.-M.}\ \bibnamefont
  {Tremblay}},\ }in\ \href {\doibase 10.1007/978-3-642-21831-6_13} {\emph
  {\bibinfo {booktitle} {Strongly Correlated Systems}}},\ \bibinfo {series}
  {Springer Series in Solid-State Sciences}, Vol.\ \bibinfo {volume} {171},\
  \bibinfo {editor} {edited by\ \bibinfo {editor} {\bibfnamefont
  {A.}~\bibnamefont {Avella}}\ and\ \bibinfo {editor} {\bibfnamefont
  {F.}~\bibnamefont {Mancini}}}\ (\bibinfo  {publisher} {Springer Berlin
  Heidelberg},\ \bibinfo {year} {2012})\ pp.\ \bibinfo {pages}
  {409--453}\BibitemShut {NoStop}%
\bibitem [{\citenamefont {Katanin}\ \emph {et~al.}(2009)\citenamefont
  {Katanin}, \citenamefont {Toschi},\ and\ \citenamefont {Held}}]{Katanin09}%
  \BibitemOpen
  \bibfield  {author} {\bibinfo {author} {\bibfnamefont {A.~A.}\ \bibnamefont
  {Katanin}}, \bibinfo {author} {\bibfnamefont {A.}~\bibnamefont {Toschi}}, \
  and\ \bibinfo {author} {\bibfnamefont {K.}~\bibnamefont {Held}},\ }\href
  {\doibase 10.1103/PhysRevB.80.075104} {\bibfield  {journal} {\bibinfo
  {journal} {Phys. Rev. B}\ }\textbf {\bibinfo {volume} {80}},\ \bibinfo
  {pages} {075104} (\bibinfo {year} {2009})}\BibitemShut {NoStop}%
\bibitem [{\citenamefont {Toschi}\ \emph {et~al.}(2007)\citenamefont {Toschi},
  \citenamefont {Katanin},\ and\ \citenamefont {Held}}]{Toschi07}%
  \BibitemOpen
  \bibfield  {author} {\bibinfo {author} {\bibfnamefont {A.}~\bibnamefont
  {Toschi}}, \bibinfo {author} {\bibfnamefont {A.~A.}\ \bibnamefont {Katanin}},
  \ and\ \bibinfo {author} {\bibfnamefont {K.}~\bibnamefont {Held}},\ }\href
  {\doibase 10.1103/PhysRevB.75.045118} {\bibfield  {journal} {\bibinfo
  {journal} {Physical Review B (Condensed Matter and Materials Physics)}\
  }\textbf {\bibinfo {volume} {75}},\ \bibinfo {eid} {045118} (\bibinfo {year}
  {2007})}\BibitemShut {NoStop}%
\bibitem [{\citenamefont {Otsuki}(2013)}]{Otsuki13}%
  \BibitemOpen
  \bibfield  {author} {\bibinfo {author} {\bibfnamefont {J.}~\bibnamefont
  {Otsuki}},\ }\href {\doibase 10.1103/PhysRevB.87.125102} {\bibfield
  {journal} {\bibinfo  {journal} {Phys. Rev. B}\ }\textbf {\bibinfo {volume}
  {87}},\ \bibinfo {pages} {125102} (\bibinfo {year} {2013})}\BibitemShut
  {NoStop}%
\bibitem [{\citenamefont {Ayral}\ and\ \citenamefont
  {Parcollet}(2015)}]{ayral15}%
  \BibitemOpen
  \bibfield  {author} {\bibinfo {author} {\bibfnamefont {T.}~\bibnamefont
  {Ayral}}\ and\ \bibinfo {author} {\bibfnamefont {O.}~\bibnamefont
  {Parcollet}},\ }\href {\doibase 10.1103/PhysRevB.92.115109} {\bibfield
  {journal} {\bibinfo  {journal} {Phys. Rev. B}\ }\textbf {\bibinfo {volume}
  {92}},\ \bibinfo {pages} {115109} (\bibinfo {year} {2015})}\BibitemShut
  {NoStop}%
\bibitem [{\citenamefont {Khurana}(1990)}]{Khurana90}%
  \BibitemOpen
  \bibfield  {author} {\bibinfo {author} {\bibfnamefont {A.}~\bibnamefont
  {Khurana}},\ }\href {\doibase 10.1103/PhysRevLett.64.1990} {\bibfield
  {journal} {\bibinfo  {journal} {Phys. Rev. Lett.}\ }\textbf {\bibinfo
  {volume} {64}},\ \bibinfo {pages} {1990} (\bibinfo {year}
  {1990})}\BibitemShut {NoStop}%
\bibitem [{\citenamefont {Hafermann}\ \emph {et~al.}(2009)\citenamefont
  {Hafermann}, \citenamefont {Li}, \citenamefont {Rubtsov}, \citenamefont
  {Katsnelson}, \citenamefont {Lichtenstein},\ and\ \citenamefont
  {Monien}}]{Hafermann09}%
  \BibitemOpen
  \bibfield  {author} {\bibinfo {author} {\bibfnamefont {H.}~\bibnamefont
  {Hafermann}}, \bibinfo {author} {\bibfnamefont {G.}~\bibnamefont {Li}},
  \bibinfo {author} {\bibfnamefont {A.~N.}\ \bibnamefont {Rubtsov}}, \bibinfo
  {author} {\bibfnamefont {M.~I.}\ \bibnamefont {Katsnelson}}, \bibinfo
  {author} {\bibfnamefont {A.~I.}\ \bibnamefont {Lichtenstein}}, \ and\
  \bibinfo {author} {\bibfnamefont {H.}~\bibnamefont {Monien}},\ }\href
  {\doibase 10.1103/PhysRevLett.102.206401} {\bibfield  {journal} {\bibinfo
  {journal} {Phys. Rev. Lett.}\ }\textbf {\bibinfo {volume} {102}},\ \bibinfo
  {pages} {206401} (\bibinfo {year} {2009})}\BibitemShut {NoStop}%
\bibitem [{\citenamefont {Hafermann}\ \emph {et~al.}(2013)\citenamefont
  {Hafermann}, \citenamefont {Werner},\ and\ \citenamefont
  {Gull}}]{Hafermann13}%
  \BibitemOpen
  \bibfield  {author} {\bibinfo {author} {\bibfnamefont {H.}~\bibnamefont
  {Hafermann}}, \bibinfo {author} {\bibfnamefont {P.}~\bibnamefont {Werner}}, \
  and\ \bibinfo {author} {\bibfnamefont {E.}~\bibnamefont {Gull}},\ }\href
  {\doibase http://dx.doi.org/10.1016/j.cpc.2012.12.013} {\bibfield  {journal}
  {\bibinfo  {journal} {Computer Physics Communications}\ }\textbf {\bibinfo
  {volume} {184}},\ \bibinfo {pages} {1280 } (\bibinfo {year}
  {2013})}\BibitemShut {NoStop}%
\bibitem [{\citenamefont {Hafermann}(2014)}]{Hafermann14}%
  \BibitemOpen
  \bibfield  {author} {\bibinfo {author} {\bibfnamefont {H.}~\bibnamefont
  {Hafermann}},\ }\href {\doibase 10.1103/PhysRevB.89.235128} {\bibfield
  {journal} {\bibinfo  {journal} {Phys. Rev. B}\ }\textbf {\bibinfo {volume}
  {89}},\ \bibinfo {pages} {235128} (\bibinfo {year} {2014})}\BibitemShut
  {NoStop}%
\bibitem [{\citenamefont {Bauer}\ \emph {et~al.}(2011)\citenamefont {Bauer},
  \citenamefont {Carr}, \citenamefont {Evertz}, \citenamefont {Feiguin},
  \citenamefont {Freire}, \citenamefont {Fuchs}, \citenamefont {Gamper},
  \citenamefont {Gukelberger}, \citenamefont {Gull}, \citenamefont {Guertler},
  \citenamefont {Hehn}, \citenamefont {Igarashi}, \citenamefont {Isakov},
  \citenamefont {Koop}, \citenamefont {Ma}, \citenamefont {Mates},
  \citenamefont {Matsuo}, \citenamefont {Parcollet}, \citenamefont
  {Pawłowski}, \citenamefont {Picon}, \citenamefont {Pollet}, \citenamefont
  {Santos}, \citenamefont {Scarola}, \citenamefont {Schollwöck}, \citenamefont
  {Silva}, \citenamefont {Surer}, \citenamefont {Todo}, \citenamefont {Trebst},
  \citenamefont {Troyer}, \citenamefont {Wall}, \citenamefont {Werner},\ and\
  \citenamefont {Wessel}}]{ALPS2}%
  \BibitemOpen
  \bibfield  {author} {\bibinfo {author} {\bibfnamefont {B.}~\bibnamefont
  {Bauer}}, \bibinfo {author} {\bibfnamefont {L.~D.}\ \bibnamefont {Carr}},
  \bibinfo {author} {\bibfnamefont {H.~G.}\ \bibnamefont {Evertz}}, \bibinfo
  {author} {\bibfnamefont {A.}~\bibnamefont {Feiguin}}, \bibinfo {author}
  {\bibfnamefont {J.}~\bibnamefont {Freire}}, \bibinfo {author} {\bibfnamefont
  {S.}~\bibnamefont {Fuchs}}, \bibinfo {author} {\bibfnamefont
  {L.}~\bibnamefont {Gamper}}, \bibinfo {author} {\bibfnamefont
  {J.}~\bibnamefont {Gukelberger}}, \bibinfo {author} {\bibfnamefont
  {E.}~\bibnamefont {Gull}}, \bibinfo {author} {\bibfnamefont {S.}~\bibnamefont
  {Guertler}}, \bibinfo {author} {\bibfnamefont {A.}~\bibnamefont {Hehn}},
  \bibinfo {author} {\bibfnamefont {R.}~\bibnamefont {Igarashi}}, \bibinfo
  {author} {\bibfnamefont {S.~V.}\ \bibnamefont {Isakov}}, \bibinfo {author}
  {\bibfnamefont {D.}~\bibnamefont {Koop}}, \bibinfo {author} {\bibfnamefont
  {P.~N.}\ \bibnamefont {Ma}}, \bibinfo {author} {\bibfnamefont
  {P.}~\bibnamefont {Mates}}, \bibinfo {author} {\bibfnamefont
  {H.}~\bibnamefont {Matsuo}}, \bibinfo {author} {\bibfnamefont
  {O.}~\bibnamefont {Parcollet}}, \bibinfo {author} {\bibfnamefont
  {G.}~\bibnamefont {Pawłowski}}, \bibinfo {author} {\bibfnamefont {J.~D.}\
  \bibnamefont {Picon}}, \bibinfo {author} {\bibfnamefont {L.}~\bibnamefont
  {Pollet}}, \bibinfo {author} {\bibfnamefont {E.}~\bibnamefont {Santos}},
  \bibinfo {author} {\bibfnamefont {V.~W.}\ \bibnamefont {Scarola}}, \bibinfo
  {author} {\bibfnamefont {U.}~\bibnamefont {Schollwöck}}, \bibinfo {author}
  {\bibfnamefont {C.}~\bibnamefont {Silva}}, \bibinfo {author} {\bibfnamefont
  {B.}~\bibnamefont {Surer}}, \bibinfo {author} {\bibfnamefont
  {S.}~\bibnamefont {Todo}}, \bibinfo {author} {\bibfnamefont {S.}~\bibnamefont
  {Trebst}}, \bibinfo {author} {\bibfnamefont {M.}~\bibnamefont {Troyer}},
  \bibinfo {author} {\bibfnamefont {M.~L.}\ \bibnamefont {Wall}}, \bibinfo
  {author} {\bibfnamefont {P.}~\bibnamefont {Werner}}, \ and\ \bibinfo {author}
  {\bibfnamefont {S.}~\bibnamefont {Wessel}},\ }\href {\doibase
  10.1088/1742-5468/2011/05/P05001} {\bibfield  {journal} {\bibinfo  {journal}
  {Journal of Statistical Mechanics: Theory and Experiment}\ }\textbf {\bibinfo
  {volume} {2011}},\ \bibinfo {pages} {P05001} (\bibinfo {year}
  {2011})}\BibitemShut {NoStop}%
\bibitem [{\citenamefont {Gull}\ and\ \citenamefont {Millis}(2012)}]{Gull12}%
  \BibitemOpen
  \bibfield  {author} {\bibinfo {author} {\bibfnamefont {E.}~\bibnamefont
  {Gull}}\ and\ \bibinfo {author} {\bibfnamefont {A.~J.}\ \bibnamefont
  {Millis}},\ }\href {\doibase 10.1103/PhysRevB.86.241106} {\bibfield
  {journal} {\bibinfo  {journal} {Phys. Rev. B}\ }\textbf {\bibinfo {volume}
  {86}},\ \bibinfo {pages} {241106} (\bibinfo {year} {2012})}\BibitemShut
  {NoStop}%
\bibitem [{\citenamefont {Galitskii}\ and\ \citenamefont
  {Migdal}(1958)}]{Galitskii58}%
  \BibitemOpen
  \bibfield  {author} {\bibinfo {author} {\bibfnamefont {V.}~\bibnamefont
  {Galitskii}}\ and\ \bibinfo {author} {\bibfnamefont {A.}~\bibnamefont
  {Migdal}},\ }\href@noop {} {\bibfield  {journal} {\bibinfo  {journal} {Zhur.
  Eksptl’. i Teoret. Fiz.}\ }\textbf {\bibinfo {volume} {34}} (\bibinfo
  {year} {1958})}\BibitemShut {NoStop}%
\bibitem [{\citenamefont {Di~Marco}\ \emph {et~al.}(2009)\citenamefont
  {Di~Marco}, \citenamefont {Min\'ar}, \citenamefont {Chadov}, \citenamefont
  {Katsnelson}, \citenamefont {Ebert},\ and\ \citenamefont
  {Lichtenstein}}]{DiMarco09}%
  \BibitemOpen
  \bibfield  {author} {\bibinfo {author} {\bibfnamefont {I.}~\bibnamefont
  {Di~Marco}}, \bibinfo {author} {\bibfnamefont {J.}~\bibnamefont {Min\'ar}},
  \bibinfo {author} {\bibfnamefont {S.}~\bibnamefont {Chadov}}, \bibinfo
  {author} {\bibfnamefont {M.~I.}\ \bibnamefont {Katsnelson}}, \bibinfo
  {author} {\bibfnamefont {H.}~\bibnamefont {Ebert}}, \ and\ \bibinfo {author}
  {\bibfnamefont {A.~I.}\ \bibnamefont {Lichtenstein}},\ }\href {\doibase
  10.1103/PhysRevB.79.115111} {\bibfield  {journal} {\bibinfo  {journal} {Phys.
  Rev. B}\ }\textbf {\bibinfo {volume} {79}},\ \bibinfo {pages} {115111}
  (\bibinfo {year} {2009})}\BibitemShut {NoStop}%
\end{thebibliography}%

\end{document}